\documentclass[12pt]{article}
\usepackage{graphicx}
\usepackage{subfigure}
\usepackage{amssymb}
\usepackage{amsfonts}
\usepackage{latexsym}
\usepackage[dvips]{color}

\input epsf.tex

\newcommand{\beq}{\begin{equation}}
\newcommand{\eeq}{\end{equation}}
\newcommand{\be}{\begin{equation}}
\newcommand{\ee}{\end{equation}}
\newcommand{\bea}{\begin{eqnarray}}
\newcommand{\eea}{\end{eqnarray}}

\makeatletter

\@addtoreset{equation}{section}
\makeatother
\newcommand{\ap}{\mbox{$\alpha^\prime$}}  

\def\href#1#2{#2}

\def\p{\partial}

\begin{document}

\baselineskip=15.5pt
\pagestyle{plain}
\setcounter{page}{1}

\begin{titlepage}
\begin{flushleft}
       \hfill                       FIT HE - 15-01 \\
       \hfill                       
\end{flushleft}

\begin{center}
  {\huge Chiral Symmetry of SYM theory in hyperbolic space  
   \vspace*{2mm}
at finite temperature \vspace*{2mm}
}
\end{center}

\begin{center}

\vspace*{5mm}
{\large ${}^{\dagger}$Kazuo Ghoroku\footnote[1]{\tt gouroku@dontaku.fit.ac.jp},
${}^{\ddagger}$Masafumi Ishihara\footnote[2]{\tt masafumi@wpi-aimr.tohoku.ac.jp},
${}^{\S}$Motoi Tachibana\footnote[3]{\tt motoi@cc.saga-u.ac.jp},\\
and ${}^{\P}$Fumihiko Toyoda\footnote[4]{\tt ftoyoda@fuk.kindai.ac.jp}
%
}\\

\vspace*{2mm}
{${}^{\dagger}$Fukuoka Institute of Technology, Wajiro, 
Higashi-ku} \\
{
Fukuoka 811-0295, Japan\\}
\vspace*{2mm}
{
${}^{\ddagger}$WPI-Advanced Institute for Materials Research (WPI-AIMR),}\\
{
Tohoku University, Sendai 980-8577, Japan\\}
\vspace*{2mm}
{${}^{\S}$Department of Physics, Saga University, Saga 840-8502, Japan\\}
\vspace*{2mm}
{
${}^{\P}$Faculty of Humanity-Oriented Science and
Engineering, Kinki University,\\ Iizuka 820-8555, Japan}

\vspace*{3mm}
\end{center}

\begin{center}
{\large Abstract}
\end{center}
We study a
holographic gauge theory living in the AdS$_4$ space-time at finite temperature.
The gravity dual is obtained as a solution of the type IIB superstring theory with two free
parameters, which correspond to four dimensional (4D) cosmological constant ($\lambda$) 
and the dark radiation ($C$) respectively. The theory studied here is in confining and chiral symmetry broken phase for $\lambda <0$ and small $C$. 
When $C$ is increased, the transition to the deconfinement phase has been observed at a finite
value of $C/|\lambda|$.
It is shown here that the chiral symmetry is still broken for a finite range of $C/|\lambda|$ in the deconfinement phase.
In other words, the chiral phase transition occurs at a larger value of $C/|\lambda|$
than that of the deconfinement transition. So there is a parameter range of a new  
deconfinement phase with broken chiral symmetry.
In order to study the properties of this phase,  
we performed a holographic analysis for the meson mass-spectrum and other quantities in terms of the probe D7 brane.
The results are compared with a linear sigma model. Furthermore, the entanglement entropy
is examined to search for a sign of the chiral phase trantion.
Several comments are given for these analyses.

\noindent

\begin{flushleft}

\end{flushleft}
\end{titlepage}

\vspace{1cm}
\section{Introduction}

The holographic approach \cite{ads1,ads2,ads3}
is expected to be applicable also to the supersymmetric Yang Mills (SYM) theory in 
curved space-time as well as in the flat Minkowski space-time. 
It would be interesting to make clear the properties of SYM theory
in the curved space-time and in cosmologically developing universe.
In this direction, some approaches have been extended to the field theory in the 
Friedmann-Robertson-Walker (FRW) type space-time \cite{H,GIN1,GIN2}.
In this case,
the 4D cosmological constant ($\lambda$) can be introduced
as a free parameter in obtaining the 5D sector of the 10D supergravity
solution. The dynamical properties of the 
4D SYM theory on the boundary are then characterized by the sign of $\lambda$. Through
the holographic approach, it has been found
that the SYM theory is in the confinement (deconfinement) phase for negative (positive) $\lambda$
\cite{H,GIN1,GIN2}. This implies that the dynamical properties of the SYM fields are largely 
influenced by the geometry of the background space-time.

Furthermore, in this approach, one more free parameter ($C$) can be introduced as an integration
constant in the solution of the supergravity. At first, this term has been 
added as the "dark radiation" to the 5D supergravity solution in the context of the brane world
\cite{BDEL,Lang}. Since it is defined in the 5D space-time, then its meaning in the 4D theory
was mysterious. In this context, afterward, it could be interpreted as
the projection of the 5D Weyl term \cite{SMS,SSM}. On the other hand,
from the viewpoint of holography, it has been found for $\lambda=0$ that $C$ corresponds to
the thermal radiation of the SYM fields at a finite temperature \cite{EGR}, then
the system is in the deconfinement phase. Actually, the 5D metric in this case
can be rewritten to the AdS$_5$-Schwarzschild form, where $C(>0)$ corresponds to the black hole mass
in this metric.
 
\begin{figure}[htbp]
\vspace{.3cm}
\begin{center}
\includegraphics[width=8cm]{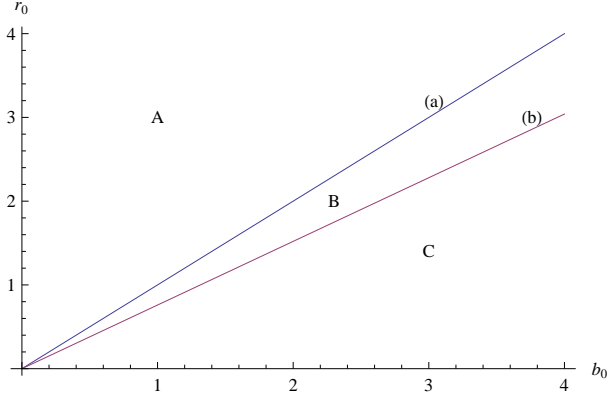}
\caption{ The line (a) shows the critical line $r_0=b_0$ separating the quark-confinement phase (A) and the deconfinement phase (B), (C). The lower critical line 
(b) $r_0=0.76 b_0$ ($\chi$-br) is separating 
the chiral symmetry broken phase, (A), (B) and the symmetry restored phase (C). }
\label{Phase-diagram}
\end{center}
\end{figure}

It is easy to imagine that the above two parameters, $C$ and the negative $\lambda$,
compete with each other to realize the opposite phase of the theory, namely 
the confinement and the deconfinement respectively. In fact,  
we find a phase transition at the point where these two opposite effects are balanced
\cite{EGR,EGR2,GN13,GNI13,GNI14}. 
As a result, the SYM theory is in the deconfinement phase for $b_0>r_0$. Here
the density of dark radiation $C$ and the 
magnitude of $|\lambda|$ are denoted by using $b_0$ and $r_0$, which
are shown in the formula (\ref{sol-12-1}) of the next section. By using these parameters,
the phase diagram of the SYM theory in the FRW space-time is obtained as in the 
Fig. \ref{Phase-diagram}. 

In the deconfinement phase, the temperature $T$, 
which is given by the Hawking temperature of the 5D metric, appears. 
The critical temperature ($T_c$) of confinement deconfinement 
transition is given
as $T_c=0$, which corresponds to the critical line $r_0=b_0$ in the Fig. \ref{Phase-diagram}. 
In the region $b_0>r_0$, the temperature monotonically increases with $b_0$ for fixed $r_0$.

It becomes possible to study the properties of the finite temperature SYM theory 
in the deconfinement phase in the FRW space-time, where the three space is hyperbolic one.
As mentioned above, for the case of $\lambda=0$ and finite $b_0$, 
the bulk solution is reduced to the well known AdS$_5$-Schwarzschild. So the SYM theory 
in its special limit of this background has been studied already, see for example \cite{GSUY}.
In this case, the two critical temperatures of 
confinement-deconfinement ($T_c$) and the chiral symmetry restoration ($T_{\chi}$) transitions 
are the same, namely $T_c=T_{\chi}=0$.
On the other hand, we show here that the value of $T_{\chi}$ shifts from $T_c=0$.
Namely we find $0=T_c<T_{\chi}$ in the case of finite $\lambda(<0)$, and then the
critical line is given by $b_0=0.76 r_0$. It is shown in the Fig. \ref{Phase-diagram}
in the parameter space of $b_0$ and $r_0$. Therefore there exists a new phase in the region
B of the Fig. \ref{Phase-diagram}, where quark and gluons
are not confined but the chiral symmetry is not yet restored.

This implies a non-trivial thermal property of the SYM theory in the FRW space-time
for negative $\lambda$. 
In order to show and understand the details of this point,
we have examined the embedding of the D7 brane in the background. For the embedded
D7 brane, we have examined
also the spectrum of the Nambu-Goldstone
boson and a massive meson modes in the region of the newly found 
deconfinement phase where the chiral symmetry is still broken.
Then we compare our results with some theoretical consequences obtaind from
some typical phenomenological models for the chiral symmetry breaking of QCD
in order to make clear the dynamical properties
implied by our holographic model.

\vspace{.3cm}
The outline of this paper is as follows.
In the next section, a 5D space-time is given by a solution of
supergravity as the dual of SYM theory in the AdS$_4$ background. In Sec.3,
the spontaneous chiral symmetry breaking is studied by embedding the D7 probe brane,
and then the spectra of two scalar mesons are examined. Since one of them corresponds
to the Nambu-Goldstone boson, then the Gell-Mann-Oakes-Renner relation and pion decay
constant are also examined. In Sec.4, the mass relations of the two scalars are discussed
in terms of the sigma model. We suggest a modified sigma model which implied by
our holographic analysis. In Sec.5, we searched a sign of the chiral transition in the
entanglement entropy. We could find a change of the behavior of the entanglement
entropy, but it could not lead to the phase transition.
The summary and discussions are given in the final section.

\section{Gravity dual}

The holographic dual to the large $N$ gauge theory embedded in 4D space-time with dark energy 
and "dark radiation" is solved by the gravity on the following form of the metric,
\beq\label{10dmetric-2-1}
ds^2_{10}={r^2 \over R^2}\left(-\bar{n}^2dt^2+\bar{A}^2a_0^2(t)\gamma_{ij}(x)dx^idx^j\right)+
\frac{R^2}{r^2} dr^2 +R^2d\Omega_5^2 \ ,
\eeq
where
\beq\label{AdS4-30} 
    \gamma_{ij}(x)=\delta_{ij} \gamma^2(x)\, , \quad \gamma (x)
  =1/\left( 1+k{\bar{r}^2\over 4\bar{r_0}^2} \right)\, , \quad 
    \bar{r}^2=\sum_{i=1}^3 (x^i)^2\, ,
\eeq
and $k=\pm 1,$ or $0$. The arbitrary scale parameter  $\bar{r_0}$ of three space
is set hereafter as $\bar{r_0}=1$.
The solution is obtained from 10D supergravity
of type IIB theory \cite{EGR,EGR2,GN13,GNI13}. 

The factors $\bar{A}$ and $\bar{n}$ are obtained by introducing two free parameters
as mentioned below. Here, we use the following form of solution,
\bea
 \bar{A}&=&\left(\left(1+\left({r_0\over r}\right)^2\right)^2+\left({b_0\over r}\right)^{4}\right)^{1/2}\, , \label{sol-10-1} \\
\bar{n}&=&{\left(1+\left({r_0\over r}\right)^2\right)^2
-\left({b_0\over r}\right)^{4}\over \bar{A}
       }\, , \label{sol-11-1}
\eea
\beq
 r_0=\sqrt{|\lambda|}R^2/2\, , \quad 
b_0=R\tilde{c}_0\, , \quad \tilde{c}_0=CR^3/(4a_0^4)\, , \label{sol-12-1}
\eeq
where the dark radiation $C$ is introduced as an integral constant in solving the equation
of motion. This solution expresses the case of negative $\lambda$.
Here, the "dark energy" 
(or cosmological term) $\lambda(t)$ is
introduced by the following equation,
\beq\label{Fried-1}
  \left({\dot{a}_0\over a_0}\right)^2+{k\over a_0^2}=\lambda\, . \label{bc-3-1}
\eeq
Although it is possible to consider a time dependent $\lambda$ as in 
\cite{EGR}, we set it here as a constant $\lambda$ for simplicity.
{In the following, our discussion would be concentrated to the case of
negative constant $\lambda$ and we assume small time derivative of $a_0(t)$. 
We should notice the following fact that the solution
$a_0=1/\sqrt{|\lambda|}=$constant is actually
allowed for negative constant $\lambda$ and $k=-1$. 

\newpage

\section{Chiral phase transition at finite temperature}\label{sec22}

\vspace{.5cm}
\subsection{D7 brane embedding}

We study the chiral condensate and the $q-\bar{q}$ meson spectrum of the boundary theory
by embedding the probe D7 brane for the flavor quarks.
The D7-brane action is given by the Dirac-Born-Infeld (DBI) and the
Chern-Simons (CS) terms as follows, 
\bea
S_{D7}&=&-\tau_{7}\int d^8\xi
 e^{-\Phi}\sqrt{-\det\left(g_{ab}+2\pi\ap F_{ab}\right)}  \nonumber \\
    && +T_{7}\int \sum_i
\left(e^{2\pi\ap F_{(2)}}\wedge c_{(a_1\ldots
a_i)}\right)_{0\ldots 7}~, \label{d7action}  \\
g_{ab}&\equiv&\p_a X^{\mu}\p_b X^{\nu}G_{\mu\nu}~, \qquad
c_{(a_1\ldots
a_i)}\,\equiv\,\p_{a_1}X^{\mu_1}\ldots\p_{a_i}X^{\mu_i}C_{\mu_1\ldots\mu_i}~, \nonumber
\eea
where $\tau_7$ is the brane tension.  
The DBI action involves
the induced metric $g_{ab}$ and the $U(1)$ world volume
field strength $F_{(2)}=d A_{(1)}$.

\vspace{.6cm}
The solution given in the previous section is obtained for the case
of constant dilaton. So it does not play any role in the present case.
For simplicity, we consider the dual theory on the boundary at $r=\infty$.
Then the induced metric for the above D7 brane is obtained as follows.
Consider the above background (\ref{10dmetric-2-1}) and rewrite it as follows
\bea
ds^2_{10}&=&{r^2 \over R^2}ds^2_{(4)}+
\frac{R^2}{r^2} dr^2 +R^2d\Omega_5^2 \ , \label{10dmetric-2-2} \\
&=&{r^2 \over R^2}ds^2_{(4)}+\frac{R^2}{r^2}\left(
       d\rho^2+\rho^2 d\Omega_3^2+\sum_{i=8}^9{dX^i}^2\right)\, , 
\eea
where
\bea
 ds^2_{(4)}&=&\left(-\bar{n}^2dt^2+\bar{A}^2a_0^2(t)\gamma_{ij}(x)dx^idx^j\right)\, , \\
 r^2&=&\rho^2+(X^8)^2+(X^9)^2\, .
\eea
Then, the induced metric of the D7 brane is obtained as
\beq\label{8dmetric}
ds^2_{8}={r^2 \over R^2}ds^2_{(4)}+\frac{R^2}{r^2} \left(
       (1+{w'}^2)d\rho^2+\rho^2 d\Omega_3^2 \right) \, , 
\label{finite-c-sol-3}
\eeq
where the profile of the D7 brane is taken as $(X^8,X^9)=(w(\rho),0)$
and $w'=\partial_{\rho}w$, then 
\beq
  r^2=\rho^2+{w}^2\, .
\eeq
In the present case,
there is no R-R filed, so the action is given as
\beq\label{D7-embed}
  S_{D7}=-\tau_{7}\Omega_3\int d^4x a_0^3(t)\gamma^3(x)\int d\rho \rho^3
   \bar{n}\bar{A}^3\sqrt{1+{w'}^2(\rho)}\, ,
\eeq 
where $\Omega_3$ denotes the volume of $S^3$ of the D7's world volume.

\vspace{.5cm}
From this action, the equation of motion for $w$ is obtained as
\beq\label{eq-p}
  w''+\left({3\over \rho}+{\rho+ww'\over r}\partial_r (\log (\bar{n}\bar{A}^3))\right)
     w'(1+{w'}^2)-{w\over r}(1+{w'}^2)^2\partial_r ( \log (\bar{n}\bar{A}^3) )=0\, .
\eeq
The constant $w\neq 0$ is not the solution of this equation, so the supersymmetry is broken
except for the case of trivial solution $w=0$.

\subsection{Embedded solutions and chiral symmetry breaking}

\vspace{.5cm}
In the confinement region, the chiral symmetry is spontaneously broken as shown in 
\cite{EGR}. The phase transition of the present model
to the deconfinement occurs when the density of the
dark radiation increases and excesses a critical point, which is given by $b_0=r_0$ 
(See the Fig. \ref{Phase-diagram}).
From this point, the Hawking temperature, 
$T_H(={\sqrt{2}b_0\over \pi R^2}\sqrt{1-(r_0/b_0)^2})$, appears for $b_0>r_0$.
At this stage, when the temperature appears,
the chiral symmetry is usually restored. In the present case, however, 
we could observe the chiral symmetry restoration after transfering into the deconfinement
phase at a finite $T_H$, 
which depends on $r_0$, as shown in the Fig.\ref{Phase-diagram} above. 
These facts are explained below through the numerical analysis.

\vspace{.3cm}
All solutions of the above equation (\ref{eq-p}) have the following asymptotic form
\beq\label{chiral-br}
  w(\rho)=m_q+{c+4m_q r_0^2\log (\rho)\over \rho^2}+\cdots\, ,
\eeq
at large $\rho$. 
In the second term of the right hand side of (\ref{chiral-br}), 
the term proportional to $\log (\rho)$ arises from the loop corrections of the
SYM theory since the conformal symmetry would be broken 
due to the existence of the cosmological constant in the 
present case \cite{H,GIN1,GIN2}. We could show however that this term is
proportional to the quark mass $m_q$, which is given by the asymptotic value of $w(\infty)$.
In order to see the spontaneous chiral symmetry breaking, it is enough to see
chiral condensate $\langle\bar{\Psi}\Psi\rangle=c=\rho^2 w|_{\rho\to\infty}$.
Then the analysis is simply performed for $m_q=0$.

\begin{figure}[htbp]
\vspace{.3cm}
\begin{center}
\includegraphics[width=7.0cm,height=6cm]{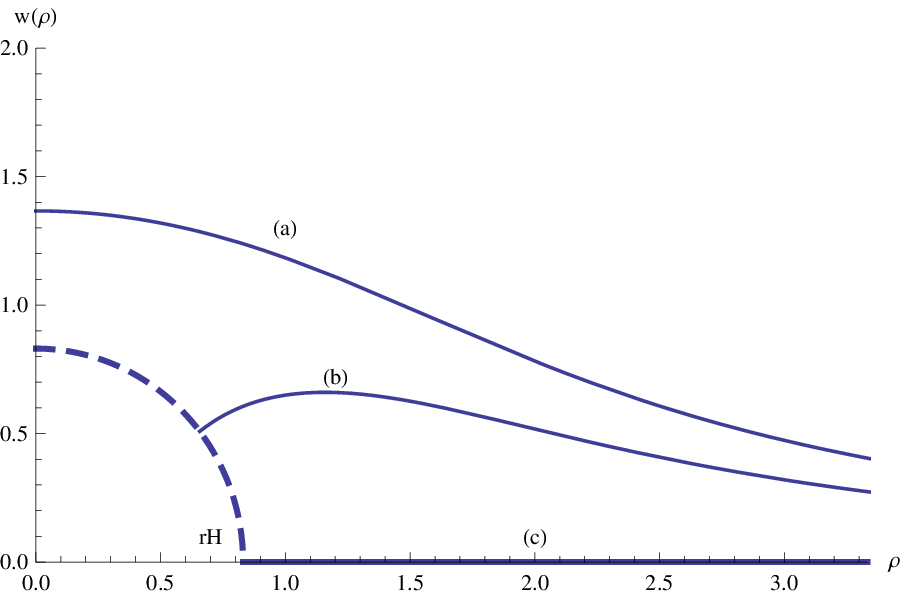}
\includegraphics[width=7.0cm,height=6cm]{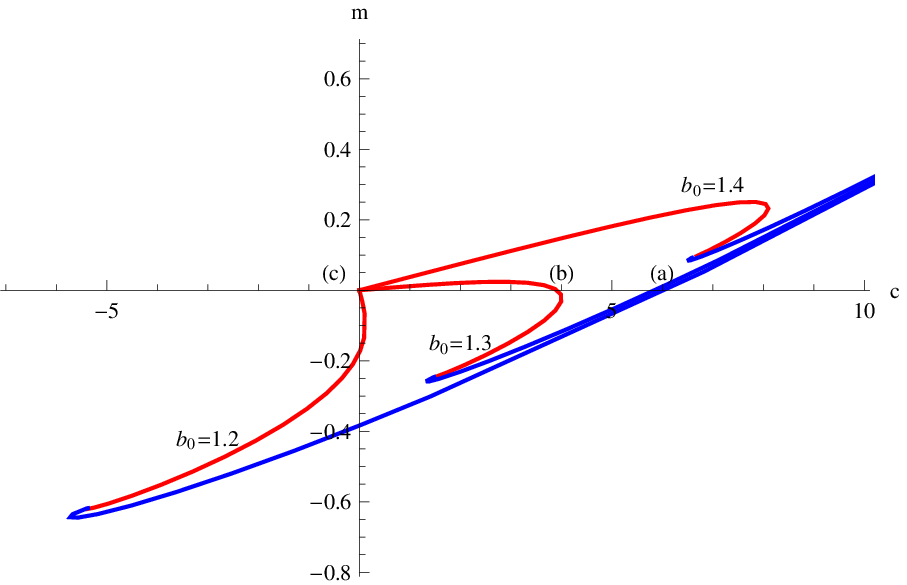}
\caption{{\bf Left;} The solutions of $w(\rho)$ with  $m_q=0$ for $b_0=1.3r_0$,  
and $r_0=1.0$. The solution (c) represents the trivial one, $w=0$.
The dashed curve denotes the horizon $r_H=\sqrt{b_0^2-r_0^2}(=0.83)$.
{\bf Right;} The $c-m_q$ relations of embedded solutions for $b_0=1.2,1.3$, and $1.4$ are
shown with the same other parameters of the left figure. For the case of $b_0=1.3$,
the embedded solutions at the three points, (a), (b), and (c), are shown in the left figure.}
\label{chiral-c2}
\end{center}
\end{figure}

We have therefore examined the numerical solutions for $m_q=0$ at various points in the parameter
space $b_0-r_0$ in order to find the transition points.
Three typical solutions are shown in the left figure of Fig.  \ref{chiral-c2}. 
They are classified as Minkowski type (M-type) ($w(0)>r_H=\sqrt{b_0^2-r_0^2}$), 
Black-hole type (B-type),
which ends on $r_H$ at $\rho=\rho_{min}$, and the trivial solution $w=0$, which is always
the solution of Eq.(\ref{eq-p}). 

We notice that the above M and B types of embedded configuration of D7 brane correspond to
the one of connected and disconnected D8 and $\bar{D8}$ branes in the Sakai-Sugimoto
model at finite temperature \cite{SS}.

We performed the numerical analyses for fixed $r_0=1$ by varying $b_0$.
In this case, the region of $b_0$ is separated to the following three one. 
\begin{itemize}
\item For $b_0>1.31$, we find 
only the trivial solution for $m_q=0$. Then the chiral symmetry is restored. 
\item For $1.28<b_0<1.31$, there are three types of solution mentioned above. 
\item For $b_0<1.28$, there are M-type and the trivial solutions. 
\end{itemize}
The distributions of the general solutions including these three types are 
shown in the right of the Fig.  \ref{chiral-c2} for $b_0=1.2,~1.3,~{\rm and} ~1.4$.
Then one might consider that
the chiral symmetry may be broken in the region $b_0<1.31$, however we must compare
the free energies in order to see which solution is favored for the given $b_0$ when plural
solutions for the same $m_q$ exist.

The free energy for each solution is obtained by substituting the solution $w(\rho)$ into the
Wick rotated Eucledian D7 action (\ref{D7-embed}). Here the normalized
free energy,
\beq
  E=\int d\rho \rho^3 \bar{n}\bar{A}^3\sqrt{1+{w'}^2(\rho)}\, ,
\eeq
is evaluated, and then the values subtracted the one for the trivial solution are shown
in the Fig. \ref{Free-energy}. 

\begin{figure}[htbp]
\vspace{.3cm}
\begin{center}
\includegraphics[width=7.0cm,height=6cm]{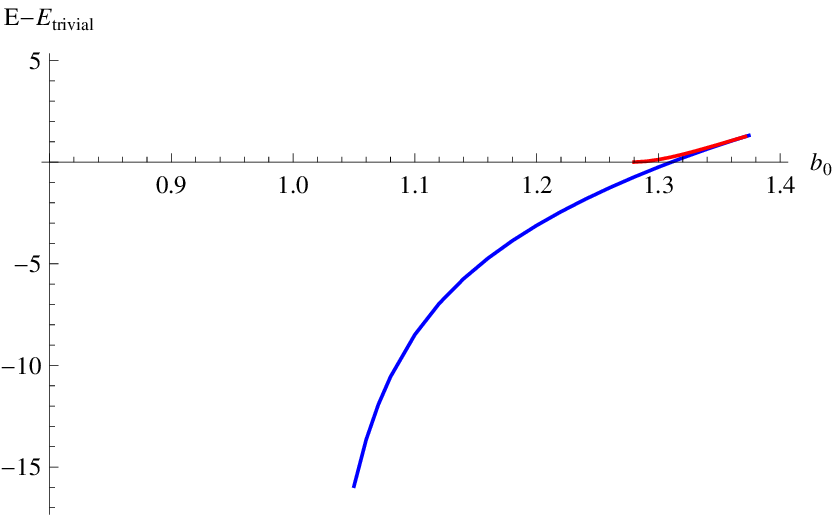}
\includegraphics[width=7.0cm,height=6cm]{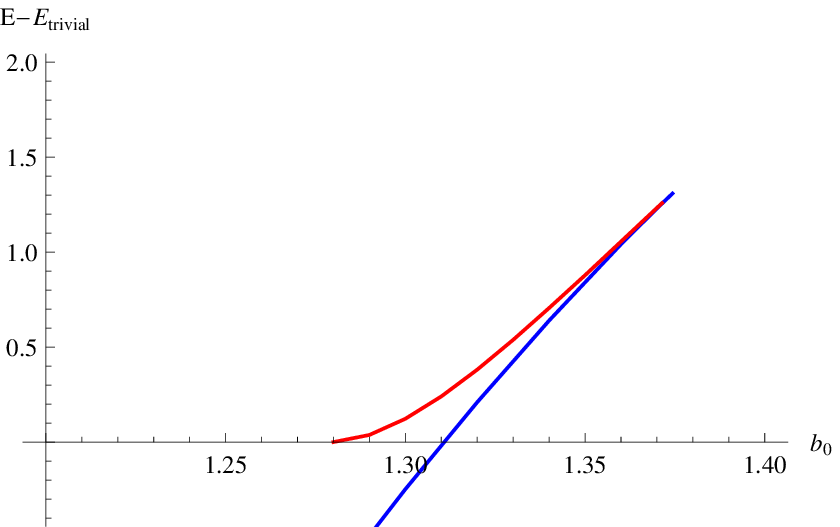}
\caption{{\bf Left;} The blue (red)  curve 
denotes the free energy difference, $E-E_{\rm trivial}$, between the one for the Minkowski 
(Black Hole) embedding
solutions of $w(\rho)$ with  $m_q=0$ and for the trivial solution $w=0$.
{\bf Right;} The part, where the blue curve crosses zero, of the left hand figure is enlarged.}
\label{Free-energy}
\end{center}
\end{figure}
This figure shows the followings. 
\begin{itemize}
\item The free energy of B-type solution is always larger than the other two. Then  this
type of solution can not be realized.
\item The value of $E_{M-type}-E_{\rm trivial}$ crosses zero at $b_0=1.31$. This implies that
a transition from M-type solution ($\langle\bar{\Psi}\Psi\rangle=c>0$) to the trivial 
solution ($\langle\bar{\Psi}\Psi\rangle=0$) occurs at this point. This point is therefore
the chiral phase transition point. The order parameter $\langle\bar{\Psi}\Psi\rangle$
has a gap at this transition point. 
\end{itemize}
The transition line  $b_0=1.31~r_0$ is shown in the Fig. \ref{Phase-diagram}.
As a result, we find two critical lines (a) and (b) in the parameter plane $(b_0,r_0)$.
The line (a) represents the transition point from confinement to deconfinement
and (b) does represent the critical line from chiral symmetry breaking phase 
to the restoring phase. 
This result implies 
the fact that the density of the dark radiation necessary for the restoration
of the chiral symmetry is larger than the one needed for realizing the deconfinement phase. 

\vspace{.5cm}
We remember that 
the role of the dark radiation is to screen the force needed for the confinement. The same 
kind of force would be necessary for the spontaneous mass generation of massless quarks. The
above result, that the chiral transition needs larger value of $C$ than the case of the confinement-
deconfinement transition, 
implies that the range of the force necessary to break the chiral symmetry
is shorter than the one for the confinement.

\section{Chiral phase transition \\ and the Nambu-Goldstone (NG) bosons}

We study the meson spectra by solving the equations of motion for the fluctuation 
of fields on the D7 brane. In the A phase in Fig.1,
we find the Nambu-Goldstone (NG) bosons and the Gell-Mann-Oakes-Renner (GOR)
relation \cite{GOR} for massless and small mass quarks, respectively. While in the B phase,
in spite of the fact that quarks are deconfined, 
we expect the existence of the NG boson and a massive scalar mode due to
spontaneous chiral symmetry breaking. 

\vspace{.5cm}

To compute the meson spectra, let us consider the fluctuations of the scalar fields 
defined as
$$X^8=w(\rho)+\tilde\phi^8, \quad X^9=\tilde\phi^9.$$
$\tilde\phi^8$ and $\tilde\phi^9$ correspond to the 
NG boson and a massive scalar boson, respectively. Note here that this phenomenon
also represents the breaking of a global $U(1)$ symmetry.

The wave functions are given in the following factorized form
$$\tilde\phi^k=\varphi^k(t,x^i)\phi_l^k(\rho){\cal Y}_l(S^3),\qquad(k=8,9)$$
where ${\cal Y}_l(S^3)$ denotes the spherical harmonic function on 
three dimensional sphere with the angular momentum $l$.  

\vspace{.3cm}
Then the linearized field equations of $\phi_l^9(\rho)$ 
and $\phi_l^8(\rho)$ for $w\neq 0$ are given as follows 
\beq
 \partial_{\rho}^2\phi_l^9+
    {1\over L_0}\partial_{\rho}(L_0)\partial_{\rho}\phi_l^9
+(1+w'~^2)
\left[({R\over r})^4{m_9^2\over \bar{n}^2}-{l(l+2)\over \rho^2}-2K_{(1)}
\right]\phi_l^9=0\, , \label{phi9}
\eeq
\beq
 L_0=\rho^3\bar{n}\bar{A}^3 {1\over\sqrt{1+w'~^2}},
  \quad    K_{(1)}={1\over \bar{n}\bar{A}^3}\partial_{r^2}(\bar{n}\bar{A}^3)
\eeq
and
$$
 \partial_{\rho}^2\phi_l^8+
    {1\over L_1}\partial_{\rho}(L_1)\partial_{\rho}\phi_l^8
+(1+w'~^2)\left[({R\over r})^4{m_8^2\over \bar{n}^2}-{l(l+2)\over \rho^2}
     -2(1+w'~^2)(K_{(1)}+2w^2K_{(2)})
\right]\phi_l^8
$$
\beq
 =-2{1\over L_1}\partial_{\rho}(L_0 w~w'K_{(1)})\phi_l^8  \label{phi8}
\eeq
\beq
 L_1={L_0\over {1+w'~^2}}, \quad
   K_{(2)}= {1\over \bar{n}\bar{A}^3}\partial_{r^2}^2(\bar{n}\bar{A}^3)\, .
\eeq

\vspace{.5cm}
Here are some remarks. 

i) 
For the 4D part of the wave-function, $\varphi^k(x^{\mu})$,  
we have assumed the following eigenvalue equation,
\beq\label{4dads}
   \square_4\varphi^k(x^{\mu})={1\over \sqrt{\tilde{g}_4}}\partial_{\mu}\sqrt{\tilde{g}_4}\tilde{g}^{\mu\nu}
   \partial_{\nu}\varphi^k(x^{\mu})=m_k^2\varphi^k(x^{\mu})
\eeq
where ${\tilde{g}_4}=-{\rm det}\tilde{g}_{\mu\nu}$ and
\beq
   \tilde{g}_{\mu\nu}dx^{\mu}dx^{\nu} = 
  \left(-dt^2+a_0^2(t)\gamma_{ij}(x)dx^idx^j\right)\, . \label{4dim-metric}
\eeq

ii) The operator $\square_4$ is derived from the eight dimensional
Laplacian $\square_8$ for the metric (\ref{finite-c-sol-3}) induced on the D7 brane.
In fact, $\square_8$ is expanded as 
\bea
 \square_8&=&{1\over \sqrt{g_{(8)}}}\partial_ag^{ab}\sqrt{g_{(8)}}\partial_b\, , \nonumber \\
       &=&{1\over \sqrt{g_{(8)}}}\partial_{\mu}g^{\mu\nu}\sqrt{g_{(8)}}\partial_{\nu}+\cdots\, ,
       \label{Laplace8}
\eea
\beq
      \sqrt{g_{(8)}}=\sqrt{g_{(4)}}\rho^3\sqrt{1+w'~^2}\, , \quad
       \sqrt{g_{(4)}}=(a_0\gamma\bar{A})^3\bar{n}\, , 
\eeq
where $a,b=0\sim 7$, $\mu,\nu=0\sim 3$ and the elliptics denotes the derivative terms with respect
to the other coordinates, $\rho$ and the one of $S^3$. 

By using the above expansion and the approximation to neglect the time derivative of $a_0(t)$, we arrive at the above equations (\ref{phi9}) and (\ref{phi8}), which are used to find the meson spectra \cite{GIN2}.

\vspace{1cm}
\subsection{Numerical Results}

Fig. 4 is the numerical results for the quark mass dependence of
the mass eigenvalues $m_8(m_s)$ and $m_9(m_{NG})$,
which correspond to $\phi^8$(massive scalar) and $\phi^9$(Nambu-Goldstone boson), respectively.
From the left panel of Fig. 4, we find the NG boson mass behaves as $m_{NG}^2\propto m_q$,
which is consistent with the GOR relation.
\par

\begin{figure}[htbp]
\vspace{.1cm}
\begin{center}
\includegraphics[width=8cm]{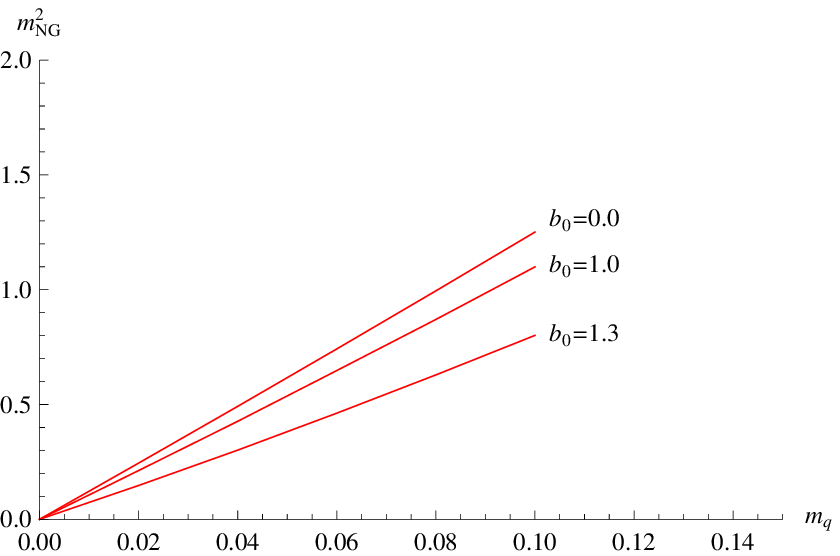}
\includegraphics[width=7cm]{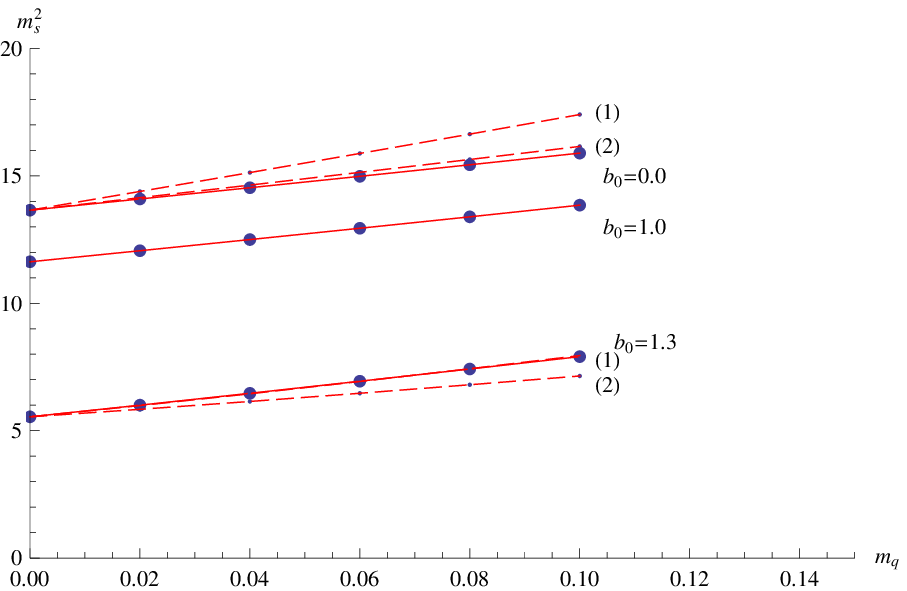}
\caption{{\bf Left:}The quark mass($m_q$) dependence of  $m_{NG}^2$ (Nambu-Goldstone boson ($\phi^9$)).
{\bf Right:} The $m_s^2$( red line) for massive scalar($\phi^8$) is shown by the solid lines. The dotted lines are the prediction from the sigma model   
given below. The line (1) denotes the one for (\ref{mass-relation}) and (2) for (\ref{mass-re-2}).}
\label{fig-mass}
\end{center}
\end{figure}

The GOR relation is expressed as
\beq\label{GOR}
m_{NG}^2= {2m_q\langle\bar\Psi \Psi\rangle\over f_{\pi}^2}\,,
\eeq
where $f_{\pi}$ denotes the pion decay constant. Since $\langle\bar\Psi \Psi\rangle$ 
is obtained for various $b_0$s through the solution of the D7 brane profile, then we can 
determine the $b_0$ dependence of $f_{\pi}$ through the GOR relation (\ref{GOR}). 
Fig. \ref{Fpi} is the numerical results.

\begin{figure}[htbp]
\vspace{.3cm}
\begin{center}
\includegraphics[width=7cm]{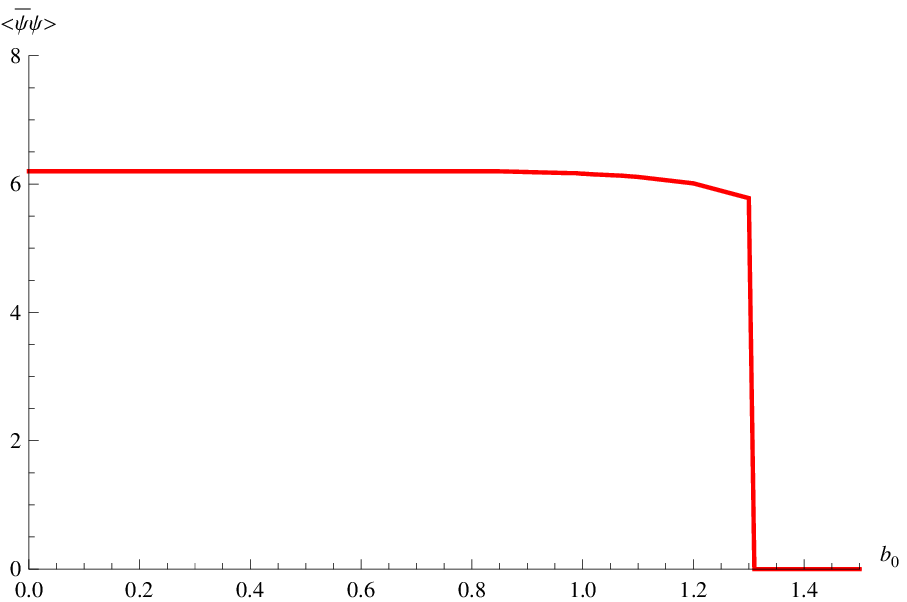}
\includegraphics[width=8cm]{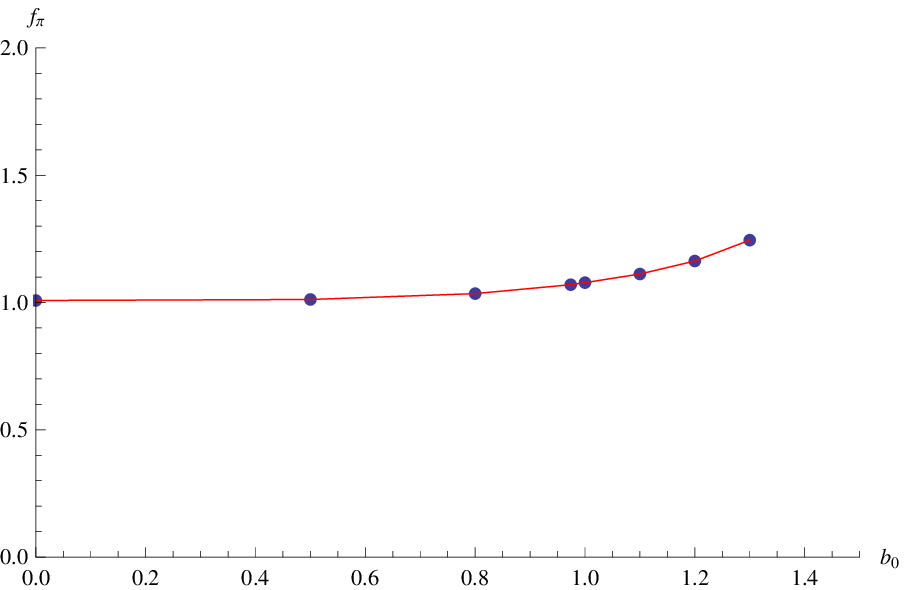}
\caption{ $b_0$ dependence of the chiral condensate $\langle\bar\Psi \Psi\rangle$ and the pion decay constant $f_{\pi}$ }
\label{Fpi}
\end{center}
\end{figure}

Here is our findings. $f_{\pi}$ is almost constant within the confinement regime
($b_0<r_0$), and then it increases with $b_0$ in the deconfinement regime ($b_0>r_0$). 
This result seems to be reasonable
since the decay channels may increase in the deconfinement regime.

\vspace{.3cm}
\subsection{Comparison with the linear $\sigma$ model} 

In principle, it would be possible to derive an effective theory of mesons
from the D7 brane action as a functional of $\phi^8$ and $\phi^9$. Then the correspondence
of the parameters in the D7 brane action to the one of the sigma model
will be obtained. However, there are various possibilities for higher order terms of the meson fields,.
Then the resultant effective action would get very complicated, which is not useful to analyze.

\vspace{.5cm}
Instead let us compare the results we obtained in the above with those from the linear sigma model. 
At the mean field level, the Lagrangian density of the linear sigma model
is given by
\begin{equation}\label{lsm}
{\cal L}=\frac{\mu^2}{2}(\sigma^2+\pi^2)-\frac{\lambda}{4}(\sigma^2+\pi^2)^2+h\sigma,
\end{equation}
where $\mu$, $\lambda$ and $h$ are the parameters while $\sigma$ and $\pi$ are the mean fields.
The last term proportional to $h$, which plays a role of the quark mass term, breaks $U(1)$ chiral symmetry explicitly.
For the small value of $h$, the vacuum is determined from the stationary conditions 
$\frac{\partial {\cal L}}{\partial \sigma}=\frac{\partial {\cal L}}{\partial \pi}=0$, which lead us to
\begin{equation}\label{lsm-vacuum}
(\sigma, \ \pi) = \left (f_{\pi}+\frac{h}{2\mu^2}, \ 0 \right )
\end{equation}
with $f_{\pi}\equiv \sqrt{\frac{\mu^2}{\lambda}}$. Then one obtains the  mass spectra such that
\begin{equation}\label{mass-relation}
M^2_{\pi}(h)=\frac{h}{f_{\pi}}, \qquad \quad M^2_{\sigma}(h)=2\mu^2+\frac{3h}{f_{\pi}}=2\mu^2+3M_{\pi}^2(h)
\end{equation}

\vspace{.5cm}
In the right panel of Fig.4, 
we compared the results (\ref{mass-relation}) with those obtained from the holographic
method. Note here that the parameter region for $b_0$ is restricted to the confinement phase
($b_0<r_0$). Hereafter, we set $r_0=1$.
The reason why we consider the linear sigma model only for the confinement 
phase is obvious because the Lagrangian density is solely written in terms of mesons.

The results (\ref{mass-relation}) definitely deviates from our holographic ones. 
It rather looks to have a good fit to the result with $b_0=1.3$. 
So we might modify the sigma model
in order to obtain a better fit to our results in the region of $b_0<1$.

\vspace{.3cm}
This is performed by adding the next order term of $(\sigma^2+\pi^2)$ as follows:
\begin{equation}\label{lsm-2}
{\cal L}=\frac{\mu^2}{2}(\sigma^2+\pi^2)-\frac{\lambda_1}{4}(\sigma^2+\pi^2)^2
   -\frac{\lambda_2}{6}(\sigma^2+\pi^2)^3+h\sigma.
\end{equation}
This type of modification is justified as far as it is based on
the derivation of the effective sigma model by expanding the D7 brane action in terms
of the $\pi$ and $\sigma$ fields. 

The vacuum is determined as the configuration $(\sigma, \ \pi) = \left (\sigma_0, \ 0 \right )$,
which is the real solution of the following equation:
\beq
   h=(-\mu^2+\lambda_1\sigma_0^2+\lambda_2\sigma_0^4)\sigma_0\, .
\eeq
Then the meson masses are given as
\begin{equation}\label{lscalar}
M^2_{\pi}=-\mu^2+\lambda_1\sigma_0^2+\lambda_2\sigma_0^4={h\over\sigma_0}, \qquad \quad 
M^2_{\sigma}=-\mu^2+3\lambda_1\sigma_0^2+5\lambda_2\sigma_0^4\, .
\end{equation}
From (\ref{lscalar}), we find 
\beq\label{mass-re-2}
   M^2_{\sigma}=2 M^2_{\pi}+\mu^2\
\eeq
by setting the following relation among the parameters,
\beq
   \lambda_2=-{\lambda_1\over 3\sigma_0^2}\, .
\eeq
This relation is a sort of tuning to obtain (\ref{mass-re-2}), which provides a better fit to
our holographic results (see the right panel of Fig. \ref{fig-mass}). 
%

\vspace{.3cm}
\section{Entanglement Entropy near the transition region}

We study the entanglement entropy ($S_{EE}$) near the transition region to find a signature of the phase transition.
$S_{EE}$ is given by calculating the minimum area of the surface $A$ whose boundary $\partial A$ is set at the boundary of the bulk. As given in \cite{RT,RT2}
the holographic entanglement entropy is given by 
\begin{equation}\label{see}
S_{EE}=\frac{S_{area}}{4G_N^{(5)}} ,
\end{equation}
where $S_{area}$ denotes the minimal surface whose boundary is defined by $\partial A$ and the surface is extended into the bulk. 

\begin{figure}[htbp]
\vspace{.3cm}
\begin{center}
\includegraphics[width=7cm]{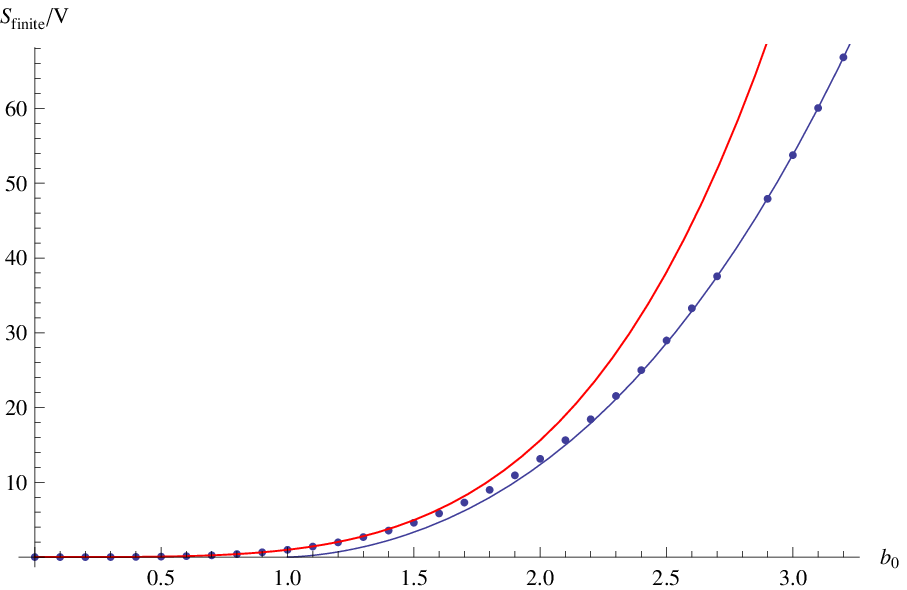}
\includegraphics[width=7cm]{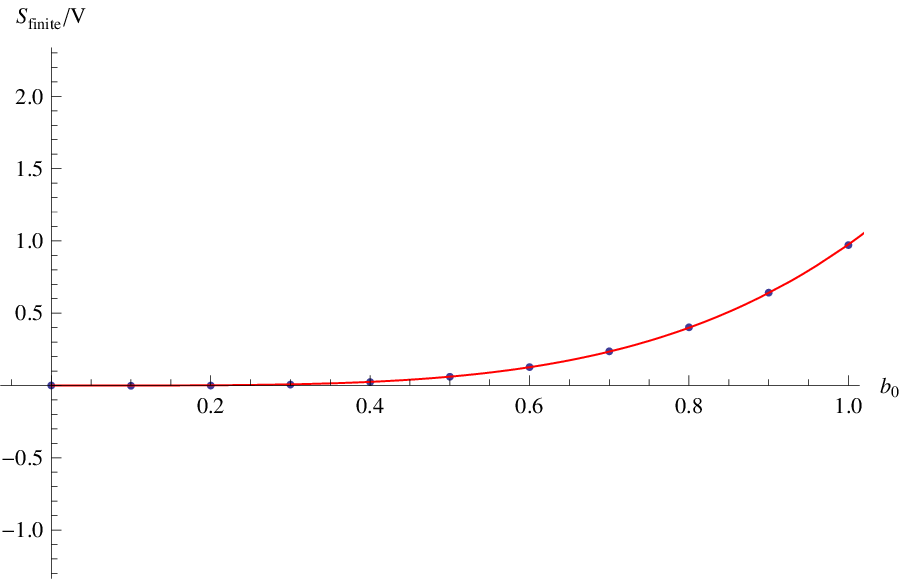}
\caption{The left figure is the relation between $S_{finite}/V$ and $b_0$ for  $r_0=R=1$ and $a_0=0.5$.  For large $b_0$ region and small $b_0$ region,  $S_{finite}/V$ can be fitted by $S_{finite}/V=26.1T_H^3$ (blue line) and  $S_{finite}/V=0.97b_0^4$(red line) respectively.
The right  figure shows the relation at small $b_0$ region.  There is no specific phenomenon at the chiral transition point( $b_0=1.31$)  }
\label{entangle}
\end{center}
\end{figure}

 We see the regularized finite part $\bar{S}_{finite}$. The detailed calculations are given in \cite{GNI13,GNI14}. This quantity
contains two contributions from the curvature $r_0$ and the dark radiation $b_0$. In order to see
how the dark radiation affects on the entropy, we consider the following quantity, 
\begin{equation}
S_{finite}\equiv \bar{S}_{finite}-\bar{S}_{finite}|_{b_0=0}\, ,
\end{equation}
by subtracting the $\bar{S}_{finite}|_{b_0=0}$ from $\bar{S}_{finite}$.

Fig. \ref{entangle} shows the relation between $S_{finite}/V$ and $b_0$. 
$V$ is the  volume of the sphere with radius $p=p_0$ in $FRW_4$ space, 
\begin{equation} \label{vol}
V=4\pi a_0^3\int_0^{p_0}\gamma^3p^2dp=2\pi a_0^3\left(\frac{4p_0(4+p_0^2)}{(p_0^2-4)^2}+\log\frac{2-p_0}{2+p_0}\right) .
\end{equation}

From the results shown in the Fig. \ref{entangle}, we can observe the followings.
\begin{itemize}
\item 
As shown in the Fig. \ref{entangle}, for small $b_0$ region, $S_{finite}/V$ is small and 
$S_{finite}/V\propto b_0^4$. The energy density of the dark radiation is also 
proportional to $b_0^4$. Thus the small deviation of $S_{finite}$ and the energy density 
will lead to the first law of the thermodynamics. On this point, we will discuss more
in the near future. 
\item For large $b_0>1$ region, on the other hand, it increases with $b_0$
rapidly and
$S_{finite}/V\propto T_H^3$ where the
Hawking temperature $T_H(b_0)$ is given as \cite{GNI14},
\beq \label{Th}
   T_H(b_0)= {\sqrt{2}b_0\over \pi R^2}\sqrt{1-(r_0/b_0)^2}\, .
\eeq
Thus, at large $b_0$ region, the entanglement entropy becomes the thermal entropy.
This behavior is expected as a high temperature behavior of the entanglement entropy $S_{finite}/V$.
\item However, as far as we observe the Fig. \ref{entangle}, there is no specific signature at the chiral transition point ($b_0\sim 1.31$) in the behavior of the entanglement entropy.
\end{itemize}

\section{Summary and Discussion}

We have studied SYM theory in the AdS$_4$ space-time. The holographic dual is expressed by a 10D supergravity solution which is described by two free parameters corresponding to the  negative 4D cosmological
constant and the dark radiation. These two quantities work to opposite direction to realize a typical phase of
the theory. The negative $\lambda$ leads the theory to the confinement phase. However the dark radiation 
prevents it. Then we find confinement-deconfinement transition at their balanced point, $r_0=b_0$, as shown previously.

Here we have pointed out that the chiral symmetry restoration does not occur yet at  $r_0=b_0$ as expected in the case of
usual QCD. The chiral transition is found at   $b_0=1.31 r_0$ after the deconfinement transition. So there exists
a new phase, where chiral symmetry is broken but the quarks and gluons are deconfined. It is shown as the region B in the phase diagram 
in the Fig.  \ref{Phase-diagram}. 
In this region, the Nambu-Goldstone boson is certenly observed, and then we could
examine for the mass spectra of mesons made of massive quark and anti-quarks to assure the GOR relation. We could study the mass
relation of the NG boson and the massive scalar modes as
expected from the sigma model, which describes well the spontaneous chiral symmetry breaking of QCD.
While the modified sigma model might be consistent with our holographic results in the confinement region A, we could not find a simple sigma model which 
is consistent in both the regions A and B.

Finally, we have examined the entanglement entropy to see the role of the dark radiation in the phase transition. In order to make clear the contribution of $b_0$, the deviation of the entanglement entropy given at $b_0>0$ from the one at $b_0=0$
is numerically studied. For small $b_0(<r_0)$, the deviation is small and increases as $b_0^4$, which is proportional to
the energy density of the dark radiation. In the large $b_0(>>r_0)$ region, it increases rapidly and it is proportional to  
$b_0^3\propto T_H^3$. This behavior is the usual thermal behavior expected in the infrared limit of the theory.
However a sharp signature of the phase transition has not been observed.

We would like to mention about relations between our results and those from some
4 dimensional effective theory.
The Polyakov-Nambu-Jona-Lasinio (PNJL) model has been often used to obtain the QCD phase diagram 
 at finite temperature/chemical potential. In this model, in addition to chiral symmetry breaking/restoration due to the quark-antiquark condensate which was originally developed in the 
Nambu-Jona-Lasinio model, the confinement-deconfinement  phase transition  can 
be taken into account by the Polyakov loop potential. According to the model calculation, 
$T_{\chi}$ gets higher than $T_c$. 
This is the similar result with ours.

 In the lattice QCD calculation, however, the result is the opposite. 
$T_{\chi}$ gets lower than $T_c$ \cite{lattice}. In \cite{Kyushu-Saga2}, the authors have 
tried to reproduce the same result from the PNJL model by including some extra terms, but 
they did not succeed. As the consequence, even among the 4 dimensional approaches,
there is no common picture for the properties of  QCD phase transition at finite temperature. 
Physically speaking, in the case with  $T_{\chi} \leq T_c$,
massless baryons appear in the temperature regime $T_{\chi} \leq T \leq T_c$ and
they would affect the thermodynamic quantities such as the equation of state. 

\par
On the other hand, the chiral phase transition in curved space has been recently discussed \cite{FF}.
It is suggested that the negative curvature $R<0$ (the negative cosmological constant)
 shifts the critical temperature $T_{\chi}$ to the increased one.

\vspace{.3cm}
\section*{Acknowledgments}
{The work of M. Ishihara was supported by World Premier International Research Center Initiative WPI, MEXT, Japan. The work of M.T. is supported in part by the JSPS Grant-in-Aid for Scientific Research, Grant No. 24540280.  }






\newpage
\end{document}